\documentclass[aps,prl,10pt,twocolumn,superscriptaddress,balancelastpage,showpacs,reprint]{revtex4-1}

\usepackage{centernot}
\usepackage{graphicx}
\usepackage{amsmath}
\usepackage{times}
\usepackage{amssymb}
\usepackage{mathrsfs}
\usepackage{chemarr}
\usepackage{color}
\usepackage{url}
\usepackage{version}
\usepackage[pdftex,colorlinks=true,
pdfstartview=FitV,
linkcolor= linkcolor,
citecolor= linkcolor,
urlcolor= linkcolor,
hyperindex=true,
hyperfigures=false]
{hyperref}

\definecolor{linkcolor}{rgb}{0,0,0.6} 

\newcommand{\qq}{\begin{eqnarray}}
\newcommand{\qqq}{\end{eqnarray}}

\newcommand{\p}{\partial}
\newcommand{\bfz}{\mathbf{z}}
\newcommand{\bfx}{\mathbf{x}}
\newcommand{\bfk}{\mathbf{k}}
\newcommand{\bfr}{\mathbf{r}}
\newcommand{\bfp}{\mathbf{p}}
\newcommand{\bfu}{\mathbf{u}}
\newcommand{\bfU}{\mathbf{U}}

\begin{document}

\title{The role of correlations in the collective behaviour of microswimmer suspensions}

\author{Joakim Stenhammar}
\email{joakim.stenhammar@fkem1.lu.se}
\affiliation{Division of Physical Chemistry, Lund University, P.O. Box 124, S-221 00 Lund, Sweden}

\author{Cesare Nardini}
\email{cesare.nardini@gmail.com}
\affiliation{DAMTP, Centre for Mathematical Sciences, Wilberforce Road, Cambridge, CB3 0WA, United Kingdom}
\affiliation{SUPA, School of Physics and Astronomy, The University of Edinburgh, James Clerk Maxwell Building, Peter Guthrie Tait Road, Edinburgh, EH9 3FD, United Kingdom}
\affiliation{Service de Physique de l'\'Etat Condens\'e, CNRS UMR 3680, CEA-Saclay, 91191 Gif-sur-Yvette, France}

\author{Rupert  W. Nash}
\affiliation{EPCC, James Clerk Maxwell Building, Peter Guthrie Tait Road, Edinburgh, EH9 3FD, United Kingdom}

\author{Davide Marenduzzo}
\affiliation{SUPA, School of Physics and Astronomy, The University of Edinburgh, James Clerk Maxwell Building, Peter Guthrie Tait Road, Edinburgh, EH9 3FD, United Kingdom}

\author{Alexander Morozov}
\email{alexander.morozov@ph.ed.ac.uk}
\affiliation{SUPA, School of Physics and Astronomy, The University of Edinburgh, James Clerk Maxwell Building, Peter Guthrie Tait Road, Edinburgh, EH9 3FD, United Kingdom}

\date{\today}

\begin{abstract}
In this Letter, we study the collective behaviour of a large number of self-propelled microswimmers immersed in a fluid. Using unprecedently large-scale lattice Boltzmann simulations, we reproduce the transition to bacterial turbulence. We show that, even well below the transition, swimmers move in a correlated fashion that cannot be described by a mean-field approach. We develop a novel kinetic theory that captures these correlations and is non-perturbative in the swimmer density. To provide an experimentally accessible measure of correlations, we calculate the diffusivity of passive tracers and reveal its non-trivial density dependence. The theory is in quantitative agreement with the lattice Boltzmann simulations and captures the asymmetry between pusher and puller swimmers below the transition to turbulence.
\end{abstract}

\maketitle

A suspension of particles that can extract energy from their surroundings and transform it into self-propulsion is an archetypal example of active matter~\cite{Ramaswamy2010,Marchetti2013RMP}. Such systems do not obey the principle of detailed balance~\cite{Cates2012RPP} at a single-particle level and their behaviour often differs significantly from that of passive suspensions at the same conditions~\cite{Poon2013}. Experiments on self-propelled particles like bacteria~\cite{dombrowski2004self,sokolov2007concentration,Clement2014}, sperm cells~\cite{Creppy2015}, mixtures of microtubules and molecular motors~\cite{Dogic2012}, vibrated granular rods~\cite{Kudrolli2003}, ``Quincke rollers''~\cite{Bricard2013Nature}, and ``colloidal surfers''~\cite{Palacci2013Science} reveal the existence of non-equilibrium steady states with non-zero macroscopic fluxes in these systems. One of the most striking examples is the phenomenon of ``bacterial turbulence''~\cite{Sokolov2012,Dunkel2013PRL,Clement2014,Wensink2012PNAS}, whereby a suspension of swimming bacteria at sufficient density forms a state with large-scale, coherent fluid motion. 

Previous analytical~\cite{saintillan2008instabilities,subramanian2009critical,Hohenegger2010,saintillan2013active} and numerical~\cite{saintillan2007,Wolgemuth2008,Saintillan2011Interface,Lushi2013,Subramanian2015JFM} studies recognise long-range hydrodynamic interactions between swimmers as a key ingredient of their collective motion; in the absence of external forces and torques, these interactions can be described by a dipolar field~\cite{Lauga2009RPP}. The main observation of the previous studies is that bacterial turbulence only emerges in suspensions of \emph{pushers} (dipolar swimmers that expel fluid along their long direction), and is absent for \emph{pullers} (dipolar swimmers that do the opposite). This conclusion has been corroborated by mean-field kinetic theories that consider the dynamics of a single swimmer in an average hydrodynamic field produced by other particles~\cite{saintillan2008instabilities,subramanian2009critical,Hohenegger2010,saintillan2013active,Subramanian2015JFM}.

\begin{figure}[h!] 
\begin{center}
\resizebox{!}{90mm}{\includegraphics{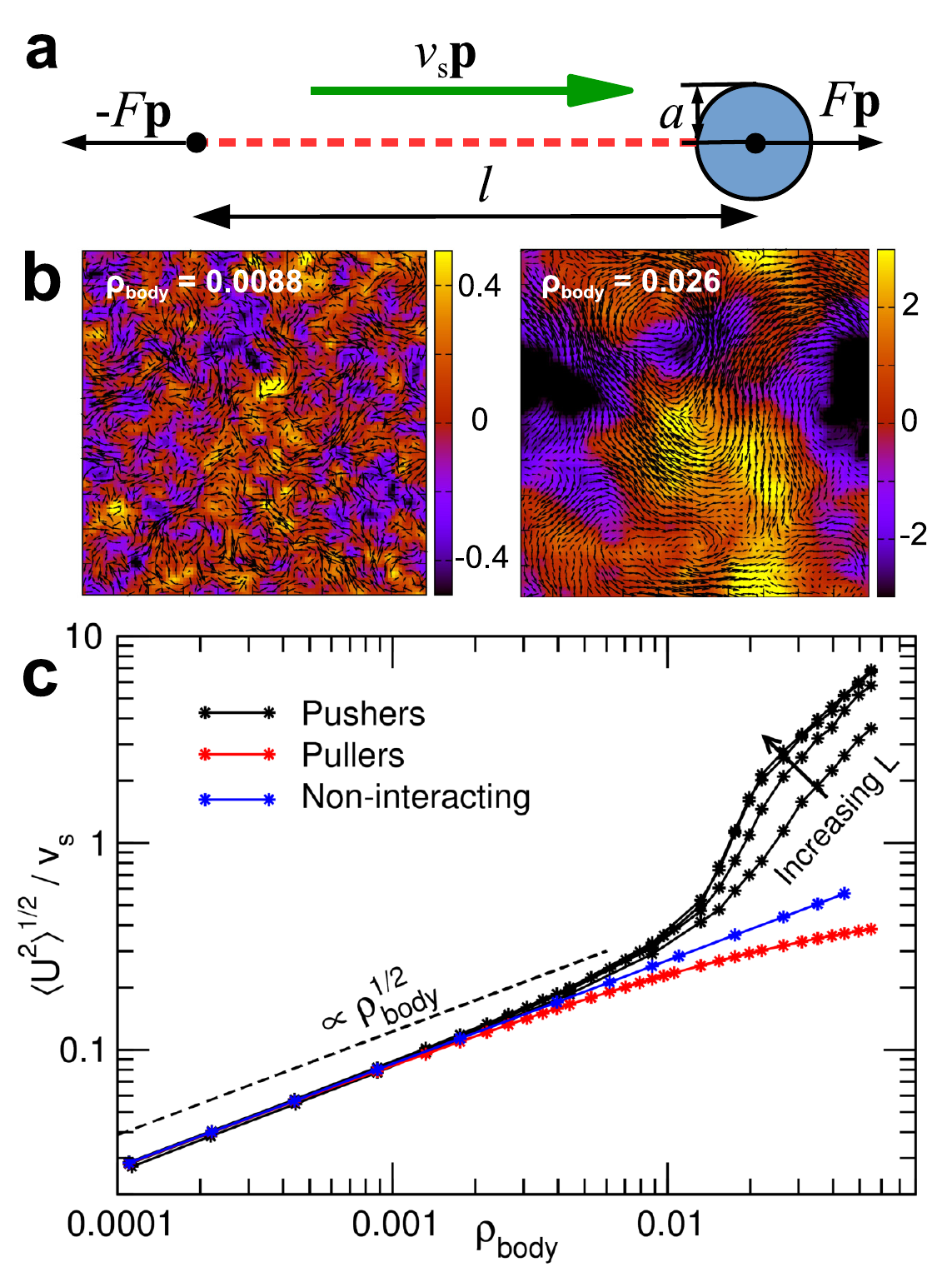}}
\caption{a) Schematic picture of the model and its parameters for a pusher swimmer. b) Snapshots showing the fluid velocity (in units of $v_s$) from LB simulations of pusher suspensions below (left) and above (right) the transition to turbulence. The vectors denote the in-plane fluid velocity, while the color map shows the out-of-plane component. c) Root-mean-square fluid velocity $\langle U^2 \rangle^{1/2}$ from LB simulations. The pusher results are presented for four different side lengths of the cubic LB box: $L = 25, 50, 100$, and $200$ in LB units (see~\cite{SI}); other curves are for $L = 100$. All densities are given in units of the swimmer body volume fraction $\rho_{\mathrm{body}} = (4\pi/3)a^3 n$, with $a \approx 0.3$ (see \cite{SI}).}
\label{fig:LBresults}
\end{center}
\end{figure}

Below the transition to collective motion, previous theoretical studies view motile suspensions as random and featureless, only acquiring non-trivial properties above the transition. In this Letter we demonstrate that due to the long-range nature of (unscreened) hydrodynamic interactions between swimmers, pre-transitional suspensions develop very strong correlations that dominate their dynamical properties (see also~\cite{Underhill2017}). Strong spatial, temporal and orientational correlations between swimmers act as precursors to bacterial turbulence and are essential to understanding the transition. We therefore develop a novel kinetic theory that goes beyond mean-field and compare its predictions to particle-resolved lattice Boltzmann (LB) simulations of up to 4 million hydrodynamically interacting microswimmers.

To illustrate the significance of swimmer-swimmer correlations, we consider the advection of passive tracer particles immersed in a microswimmer suspension. The effective tracer diffusion constant is an observable sensitive to the dynamical state of the system, and it has been extensively studied experimentally in suspensions of bacteria~\cite{Wu:00,mino2011,Poon2013PRE} and algae~\cite{Goldstein2009PRL,Polin2016NatComm,kurtuldu2011}. For low densities of swimmers, it has been predicted to scale linearly with the swimmer density and to be identical for pushers and pullers of equal dipolar strengths~\cite{childress2011,mino2011,Poon2013PRE,Yeomans2013PRL,kasyap2014hydrodynamic,morozov2014enhanced}. Here, we use LB simulations and kinetic theory to show how correlations break the pusher-puller symmetry and result in non-linear scaling of the enhanced diffusivity with swimmer density. Our analysis suggests that these correlations become significant even for densities as low as $10\%$ of the critical density, with the latter being estimated at a volume fraction of $\sim 2\%$ for \emph{E. coli}-like parameters (see below).

\emph{Model description.} We consider a 3-dimensional suspension of $N$ microswimmers immersed in a fluid of volume $V$ at number density $n = N/V$. Each swimmer exerts a force on the fluid, $-F\mathbf{p}$, representing the flagellum, and an equal and opposite force, $F\mathbf{p}$, applied a distance $l$ from the propulsive force, representing the cell body; here $\mathbf{p}$ is the swimmer orientation (see Fig. \ref{fig:LBresults}a). The body is modelled as a sphere with hydrodynamic radius $a$,  while the velocity scale of this model is defined as $v_0 \equiv F/(\mu l)$, where $\mu$ is the viscosity of the fluid. The dipolar strength of each swimmer is given as $\kappa = Fl/\mu$, with $\kappa > 0$ representing pushers and $\kappa < 0$ pullers, and the non-dimensional swimmer density is defined as $\rho = n l^3$. To facilitate comparison with experimental results, we also define the reduced density $\rho_{\mathrm{body}} = (4\pi/3) a^3 n$, where $a$ is estimated as described in~\cite{SI}, which gives an estimate of the volume fraction based on the bacterial body volume. Each swimmer $i$ moves according to the following equations of motion:
\qq
\label{rdot}\label{eq:translation} \dot{r}_i^\alpha = v_s p_i^{\alpha}+U^\alpha(\mathbf{r}_i), 
\qquad
\label{pdot}\label{eq:orientation} \dot{p}_i^\alpha = \mathbb{P}_i^{\alpha\beta} \nabla_i^\gamma U^\beta(\bfr_i) p_i^\gamma,
\qqq
where $\mathbb{P}_i^{\alpha\beta}=\delta_{\alpha\beta}- p_i^\alpha p_i^\beta$,  
$\mathbf{U}(\mathbf{r}_i)$ is the fluid velocity at the position of swimmer $i$, and $v_s$ is the swimming speed (see~\cite{SI}); Greek indices denote Cartesian components. In addition to being rotated by the fluid, the orientation of each swimmer is randomized with average tumbling frequency $\lambda$~\footnote{The rotational dynamics of Eq. \eqref{pdot} corresponds to flow-aligning swimmers, and is valid in the limit $l \gg a$; the generalisation to finite aspect ratios is straightforward and does not change the main conclusions drawn from the model.}. 

\emph{Lattice Boltzmann simulations.} Large-scale numerical simulations of up to $\mathcal{O}(10^6)$ hydrodynamically interacting microswimmers in cubic boxes with periodic boundary conditions were performed using a D3Q15 lattice Boltzmann (LB) algorithm. The swimmers are described using the point-force implementation developed by Nash \emph{et al.}~\cite{Nash2008PRE,NashPRL}, which accurately captures the full far-field interactions between the particles, while neglecting short-range hydrodynamics, lubrication effects and non-hydrodynamic interactions; see~\cite{SI} for further details. As shown in Fig. \ref{fig:LBresults}b, we qualitatively capture the transition from seemingly random motion at low density of swimmers, to bacterial turbulence at higher densities of pushers, visible as large-scale fluid vortices and jets. Quantitatively, this is characterised in Fig. \ref{fig:LBresults}c as a rapid deviation of the root-mean-square fluid velocity $\langle U^2 \rangle^{1/2}$ for pushers at $\rho_{\mathrm{body}} \approx 0.02$ (\emph{i.e.}, close to the experimentally observed transition densities \cite{Sokolov2012,Clement2014}), from the expected $\rho^{1/2}$ behaviour at low density to a state with much larger velocity fluctuations; for pullers, correlations lead to $\langle U^2 \rangle^{1/2}$ increasing slower than $\rho^{1/2}$. Figure \ref{fig:LBresults}c also highlights the strong system-size dependence of these results: velocity fluctuations change appreciably when going from a box length of $L = 25$ ($N \simeq 10^3 - 10^4$, comparable to previous particle-resolved studies~\cite{Saintillan2011Interface,Lushi2013,Subramanian2015JFM}) to $L = 200$ ($N \simeq 10^6 - 10^7$), indicating the need for very large-scale simulations (at least $L = 100$ and $N \simeq 10^5$) to describe the collective motion in these systems. We postpone further characterisation of the turbulent state to future studies, and in the following we focus on velocity fluctuations in the pre-transitional region ($\rho_{\mathrm{body}} \leq 0.02$) and the buildup of correlations that lead to bacterial turbulence. 

\begin{figure}[th] 
\begin{center}
\resizebox{!}{50mm}{\includegraphics{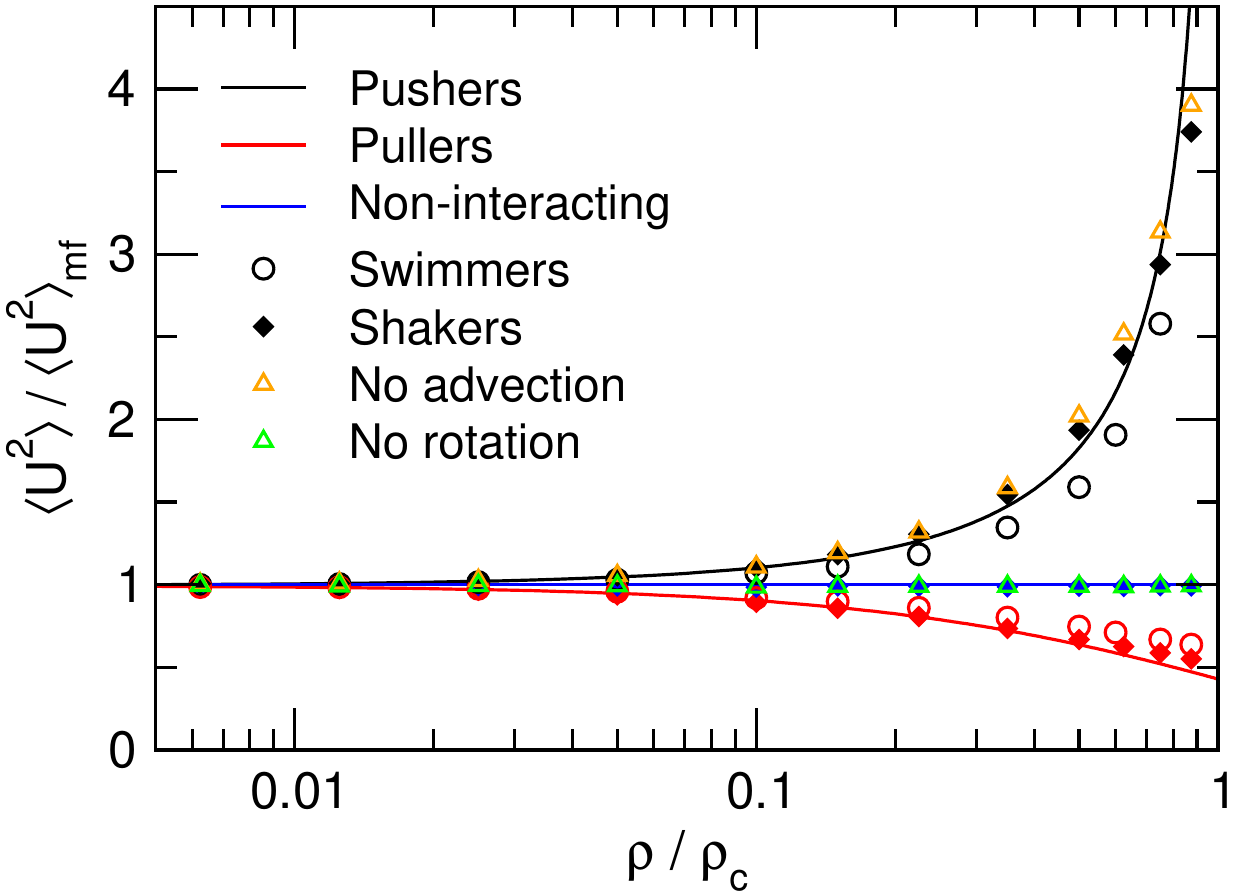}}
\caption{Variance of the fluid velocity as obtained from LB simulations (symbols) and kinetic theory (lines), normalized by its mean-field value $\langle U^2\rangle_{\mathrm{mf}} = 21v_0 \rho \kappa_n^2/(2048\epsilon_n)$. The density is normalized by the critical density $\rho_c = 5\lambda_n/|\kappa_n|$ . The blue line shows the mean-field prediction, which matches simulations when swimmer-swimmer interactions are switched off. Orange and green triangles are LB results for pusher shakers without either advection or rotation by the fluid. The theoretical results were obtained by numerically solving the full integrals given before Eq. \eqref{eq:v2:approx}}.
\label{fig:v2}
\end{center}
\end{figure}

In Fig. \ref{fig:v2}, we make the observation that, in this density regime, interactions between swimmers are dominated by their mutual rotation, while the effect of advection is secondary. There we also show results obtained for so-called \emph{shakers} -- particles that apply forces to the fluid but do not swim. These behave very similarly to swimmers, not only qualitatively~\cite{hatwalne2004rheology} but also quantitatively, indicating that the effect of swimming is subdominant. We now proceed to develop the kinetic theory, where the latter observation will prove to be mathematically convenient. 

\emph{Kinetic theory.} The central theoretical tool developed in this Letter is a kinetic theory that describes the suspension at a coarse-grained level. We assume that the swimmer density $\rho$ is sufficiently low that the single-swimmer Fourier-space velocity field $\bfu_\bfk(\bfp)$ can be described by that of a regularised point dipole~\cite{Cortez2005}:
\qq\label{eq:dipole_field}
k^2 u_\bfk^\alpha(\bfp) =
\,-i \kappa A(k\epsilon)\, 
(\bfk\cdot \bfp)\, 
[ p_\alpha - k^{-2} k_\alpha(\bfk\cdot \bfp) ].
\qqq
Here, $k=|\bfk|$, $A(x)=x^2 K_2(x)/2$ with $K_2$ the modified Bessel function of the second kind, and $\epsilon$ is a regularisation parameter. The starting point of any kinetic theory is standard \cite{balescu1975equilibrium} and is briefly summarised in the following. The dynamics of the system is described by the master equation~\cite{Gardiner:1985} for the $N$-body probability density function (PDF) $f_N(\bfz_1,...,\bfz_N,t)$, with $\bfz_i=(\bfr_i,\bfp_i)$. We then introduce the reduced PDFs $f_s=N!/(N-s)!\int d\bfz_{s+1}...d\bfz_N f_N$ and derive a BBGKY hierarchy for the $f_s$~\cite{SI}. This hierarchy is conveniently written in terms of connected correlations $g_s$ defined by $f_2(\bfz_1,\bfz_2,t)=f_1(\bfz_1,t)f_1(\bfz_2,t)+g_2(\bfz_1,\bfz_2,t)$ and similarly for higher orders.
To close the exact BBGKY hierarchy in a controlled way, we follow the approximation scheme used in equilibrium~\cite{balescu1975equilibrium,nicholson1983introduction,heyvaerts2010balescu,campa2009statistical,Bouchet2010PhysicaA} and non-equilibrium~\cite{nardini2012kinetic,nardini2012kineticlong} systems with long-range interactions, such as plasmas and self-gravitating systems. We consider the limit of a large number of swimmers, $N\gg 1$, at a \emph{fixed} density $\rho$; our approach is perturbative in the small parameter $1/N$, \emph{not} in $\rho$, and, for this reason, we get accurate predictions even close to the onset of bacterial turbulence. In this limit, and in the absence of hidden divergences \footnote{This hypothesis is checked \emph{a posteriori} showing that the final results keep the order of magnitude assumed. }, we can demonstrate that $f_1\sim \mathcal{O}(1)$ and $g_s\sim \mathcal{O}(1/N^{s-1})$. At leading order, $g_s=0$ for $s\geq 2$, we obtain the mean-field approximation already analysed in the literature for swimmers~\cite{saintillan2008instabilities,subramanian2009critical,Hohenegger2010,saintillan2013active,Subramanian2015JFM} and suspensions of passive rods~\cite{doi1988theory}. In this Letter we go beyond the mean-field approximation, retaining $g_2$ and discarding $g_s$ for $s\geq 3$~\cite{SI}; henceforth, we use $f \equiv f_1$ and $g\equiv g_2$. 

The perturbative analysis summarized above shows that $g$ solves
\qq\label{eq:Lyapunov}
\p_t g +L_f^{(1)}[g]+L_f^{(2)}[g] =  C\,,
\qqq
where~\footnote{We assume here $f=n/4\pi$, which simplifies significantly both the expressions for $C$ and for $L_f$ and corresponds to a suspension which is isotropic and homogeneous on average. Their expressions for general $f$ is given in \cite{SI}.}
\qq\label{eq:C}
C\equiv
3 \left(\frac{n}{4\pi}\right)^2 \left[
p_1^\alpha p_1^\beta \nabla^\alpha u^\beta (\bfr_1-\bfr_2,\bfp_2) \right. \nonumber \\
\left. + p_2^\alpha p_2^\beta \nabla^\alpha u^\beta (\bfr_2-\bfr_1,\bfp_1)  
\right]\,,
\qqq
and
\qq
L_f[h] &&=
\lambda h-\frac{\lambda}{4\pi}\int d\bfp' h+\nabla^\alpha (v_s p^\alpha h)\\
&&+\frac{n}{4\pi}\p^\alpha\Big( \mathbb{P}^{\alpha\beta}
\nabla^\gamma U^\beta_{mf}[h] \,p^\gamma\, 
\Big), \nonumber
\qqq
where we used the notation $\p^\alpha\equiv\mathbb{P}^{\alpha\beta}\frac{\p}{\p p^\beta}$. Here, $U_{\mathrm{mf}}^\alpha[h](\bfr) = \int d\bfr_1 d\bfp_1 \,\,u^\alpha(\bfr-\bfr_1,\bfp_1) h(\bfr_1,\bfp_1,t)$, and $L_f[h] $ is the mean-field operator $V$ linearised close to $f$ and acting on the function $h$, where $V = \nabla^\alpha(\dot{r}_{\mathrm{mf}}^\alpha f) + \p^\alpha( \dot{p}_{\mathrm{mf}}^\alpha f)+\lambda f - (\lambda/4\pi)\int d\bfp  f$ and $(\dot{\bfr}_{\mathrm{mf}},\dot{\bfp}_{\mathrm{mf}})$ are given by Eqs.~\eqref{eq:translation} with  $U^\alpha(\mathbf{r}_i)$ replaced by $U_{\mathrm{mf}}^\alpha[f](\bfr)$; $L_f^{(i)}$ acts on $\bfz_i$.

Motivated by our numerical observation that the nature of the transition and the properties of the suspension can be understood in the absence of self-propulsion, we consider only the case of  $v_s=0$. We stress that, while being significantly more complex, all the results presented below can also be obtained for $v_s > 0$, as will be shown in a forthcoming publication~\cite{NardiniSwimmersLong2017}. 

We now introduce an It\=o white noise $\eta$ with covariance $\mathbb{E}[\eta\,\eta] = C\,\delta(t-t')$, where $\mathbb{E}$ denotes the average over $\eta$, and $C$ is given by Eq. \eqref{eq:C}. Formally, the connected correlator $g$ can be written as $g(\bfz_1,\bfz_2,t)=\mathbb{E}[\delta f(\bfz_1,t)\,\delta f(\bfz_2,t)]$, where $\delta f $ solves
\qq\label{eq:kin-eq-linear-delta-f-stoch-1}
\p_t \delta f +L_f[\delta f]=\eta\,.
\qqq
Because $g$ is the covariance of density fluctuations close to $f$, these fluctuations are described by the random field $\delta f$. Being small, they are given by a linear stochastic process, Eq. \eqref{eq:kin-eq-linear-delta-f-stoch-1}, although the variance of the noise $C$ is non-trivial and could not have been guessed \emph{a priori}. It is also remarkable that, as a result of the coarse graining procedure, the density fluctuations are described by a stochastic process even when the underlying microscopic dynamics are deterministic ($\lambda=0$). Equations (\ref{eq:Lyapunov}) and (\ref{eq:kin-eq-linear-delta-f-stoch-1}) can be solved exactly to yield $\delta f = e^{-tL_f}\delta f(t=0) + \int_0^t e^{-(t-s)L_f}\eta(s)ds$ and $g = \int_0^t e^{-s L_f^{(1)}} e^{-s L_f^{(2)}} C\, ds$~\footnote{We assume here $g(t=0) = 0$. A different initial condition for $g$ only amounts to a transient, exponentially decaying contribution which does not affect the stationary properties.}.
As $\widetilde{\delta f} = e^{-tL_f}\widetilde{\delta f}(t=0)$ solves $\p_t \widetilde{\delta f} + L_f [\widetilde{\delta f} ] = 0$, to compute $g$ we only need the solution $\widetilde{\delta f}$ of the above deterministic dynamics, with appropriate initial conditions set by $C$. It turns out that $\widetilde{\delta f} $ can be found exactly for a generic initial condition~\cite{SI}. For the fluctuations of the fluid velocity in Fourier-Laplace space $\widetilde{\delta U}^{\alpha}_{\bfk}(\omega)= \int d^2\bfp'  \,u^\alpha_\bfk (\bfp')\,\widetilde{\delta f}_\bfk(\bfp',\omega) $, we then obtain the following closed expression
\qq\label{eq:vs0-deltaU-final}
\widetilde{\delta U}^{\alpha}_{\bfk}(\omega) 
 = 
 \frac{1}{C^{}_0(k,\omega)}\,
 \int d^2\bfp \frac{\widetilde{\delta f}_\bfk (\bfp,t=0)\,u_\bfk^\alpha(\bfp)}{-i\omega +\lambda}\,
\qqq
where, for the regularised dipolar field in Eq. \eqref{eq:dipole_field}, we have $C_0(k,\omega) = 1- (\kappa n/5) A(k\epsilon) / (-i\omega+\lambda)$. Equation (\ref{eq:vs0-deltaU-final}) is valid under the assumption that the dynamical state described by $f$ is linearly stable, which corresponds to the zeros $\omega_*=\omega_R+i \omega_I$ of $C_0(k,\omega)$ having negative imaginary parts in the Laplace domain, $\omega_I<0$. These zeros are given by $\omega_R=0$ and $\omega_I(k\epsilon) = -\lambda +  \kappa n A(k\epsilon)/5$, which implies that a suspension of pullers ($\kappa < 0$) is always stable, while pusher suspensions ($\kappa > 0$) are stable only if $n < 5\lambda / \kappa $, in agreement with earlier results~\cite{saintillan2008instabilities,subramanian2009critical,Hohenegger2010,saintillan2013active}. 

We note the emergence of the characteristic time-scale $\omega_I^{-1}$, which describes the typical time for a small fluctuation of the fluid velocity to relax. At $\rho = 0$ it reduces to $\omega_I^{-1} = \lambda^{-1}$, while upon increasing the density, it decreases for pullers and increases for pushers. This suggests qualitative differences between the statistical properties of suspensions of pushers and pullers even below the onset of bacterial turbulence.

\emph{Fluid velocity variance.} We now compare the kinetic theory developed above with the results of LB simulations by computing the variance $\langle U^2 \rangle$ of the fluid velocity: $\langle U^2 \rangle = \langle U^2 \rangle_{\mathrm{mf}} + \langle U^2 \rangle_{\mathrm{corr}}$, where $\langle U^2\rangle_{\mathrm{mf}} = n/(4\pi(2\pi)^3)\int d\bfk d\bfp\, u_\bfk^\alpha(\bfp) u_{-\bfk}^\alpha(\bfp)$ is the mean-field contribution and $\langle U^2 \rangle_{\mathrm{corr}}$ contains corrections induced by swimmer-swimmer correlations. Using the formal solution for $g$ and Eqs. (\ref{eq:C}) and (\ref{eq:vs0-deltaU-final}) we obtain $\langle U^2\rangle_{\mathrm{mf}} = v_0^2\frac{\rho \kappa_n^2}{15\pi^2}
 \int_0^{\infty} dk  A^{2}( \epsilon_nk)$ and $\langle U^2 \rangle_{\mathrm{corr}} = -v_0^2\frac{\rho^2 \kappa_n^3}{ 75\pi^2} \int_0^{\infty} dk  \frac{A^{2}( \epsilon_nk)}{\omega_n( \epsilon_nk)}$, which is well approximated by~\cite{SI}
\begin{align}
\label{eq:v2:approx}
\frac{\langle U^2\rangle}{v_0^2} \approx \frac{21 \rho \kappa_n^2 }{2048 \epsilon_n }\left[1\pm\frac{\rho(2\rho_c \mp \rho)}{2\rho_c (\rho_c\mp \rho)} \right],
\end{align} 
where the upper sign corresponds to pushers and the bottom one to pullers. Here, we have introduced the non-dimensional units $\kappa_n=\kappa / (l^2 v_0)$, $\epsilon_n=\epsilon/l$, $\omega_n=\omega_I l / v_0$, and $\lambda_n = \lambda l / v_0$, using $l$ and $l/v_0$ as the respective length- and time-scales. The density $\rho_c \equiv 5\lambda_n/|\kappa_n|$ corresponds to the onset of collective motion for pushers, and acts as a characteristic density scale for pullers. Fig. \ref{fig:v2} shows excellent agreement between this prediction and the LB data, even close to the onset of turbulence, emphasising the non-perturbative nature of our approach. Equation \eqref{eq:v2:approx} and the data in Fig. \ref{fig:v2} can also be used to assess the relative importance of correlations: for $\rho \lesssim 0.1\rho_c$,  pushers and pullers behave equivalently and follow the mean-field prediction, while above it, correlations have to be taken into account to obtain correct quantitative predictions. We furthermore note that, using the body volume $(4\pi/3)a^3$ with $a \approx 0.3$~\cite{SI}, we obtain $\rho_c^{\mathrm{body}} \approx 0.023$, in good agreement with the experimentally observed transitional volume fractions of $2\%$ in \emph{E. coli}~\cite{Clement2014} and \emph{B. subtilis}~\cite{Sokolov2012}.

\emph{Enhanced diffusivity}. We now consider the advection of a passive tracer with the dynamics $\dot{\bfx}=\bfU(\bfx,t)$ immersed in a suspension of shakers. Its long-time motion is diffusive \cite{Thiffeault2015}, and we use the kinetic theory developed here to calculate its effective diffusion constant $D_h$. Denoting by $\bfx_t$ the tracer position at time $t$, we have $\langle |\bfx_t-\bfx_0|^2 \rangle = 6 D_h t $, with $D_h$ related to the fluid velocity autocorrelation by~\cite{Kubo}
\qq\label{eq:tracer-diff-Kubo}
D_h =\frac{1}{3}\int_0^{\infty} ds\,\mathcal{C}(\bfx_s,s,\bfx_0,0).
\qqq
Here, $\mathcal{C}(\bfx_s,s,\bfx_0,0) = \mathbb{E}\left[ \delta U^\alpha (\bfx_s,s) \, \delta U^\alpha (\bfx_0,0)\right] $ and $\delta\bfU$ is the fluctuating fluid velocity obtained from the solution of Eq. \eqref{eq:kin-eq-linear-delta-f-stoch-1}.
The computation of $D_h$ is greatly simplified by iteratively inserting the solution of the tracer dynamics into Eq. (\ref{eq:tracer-diff-Kubo}), recalling that $\delta \bfU$ is small, and thus Taylor expanding around $\bfx_0$ (see~\cite{bouchet2005kinetics,bouchet2005anomalous}). At leading order, we obtain $D_h\simeq 
 \frac{1}{3} \int_0^{\infty} ds\, \mathcal{C}(\bfx_0,s,\bfx_0,0) $,
where corrections due to higher-order correlations have been discarded in agreement with the approximations made in the kinetic approach. Given the solution of the linear dynamics, the diffusivity reads $D_h/(l v_0) =
-\frac{ \rho \kappa_n^2}{45\pi^2 }  \int_0^{\infty} dk A^{2}( \epsilon_nk)/\omega_n(k\epsilon_n)
+\frac{\rho^2 \kappa_n^3}{225\pi^2}  \int_0^{\infty} dk A^{2}( \epsilon_nk)/\omega_n^{2}(k\epsilon_n)$, and is approximated by~\cite{SI}
\begin{align}
\label{eq:Dhapprox}
\frac{D_h}{D_h^{\mathrm{free}}} \approx \frac{1}{2}\left[ 1 + \frac{\rho_c}{\rho_c \mp \rho} \pm \left( \frac{\rho}{\rho_c}+\frac{\rho \rho_c}{(\rho_c \mp \rho)^2}\right)\right].
\end{align}
As in Eq. \eqref{eq:v2:approx}, the top (bottom) sign corresponds to pushers (pullers). In the low density limit, $D_h \to D_h^{\mathrm{free}} + \mathcal{O}(\rho^2)$, where $D_h^{\mathrm{free}}/(l v_0) = \langle U^2 \rangle_{\mathrm{mf}}/3\lambda_n v_0^2 = 7 \rho \kappa_n^2/ (2048\epsilon_n \lambda_n)$ is the diffusivity obtained from discarding interactions between shakers. An interesting observation is that $D_h^{\mathrm{free}}$ is, contrary to the variance of the fluid velocity, sensitive to the presence of self-propulsion: for shakers, $D_h^{\mathrm{free}}$ diverges as $\epsilon_n \to 0$, while an analogous computation for swimmers gives $D_h^{\mathrm{free}}/lv_0=\rho \kappa_n^2/48\pi$, in agreement with earlier theoretical predictions~\cite{childress2011,mino2011,Poon2013PRE,Yeomans2013PRL,kasyap2014hydrodynamic,morozov2014enhanced}. We have confirmed this difference of $D_h^{\mathrm{free}}$ between shakers and swimmers through LB simulations at low density.

\begin{figure}[th] 
\begin{center}
\resizebox{!}{50mm}{\includegraphics{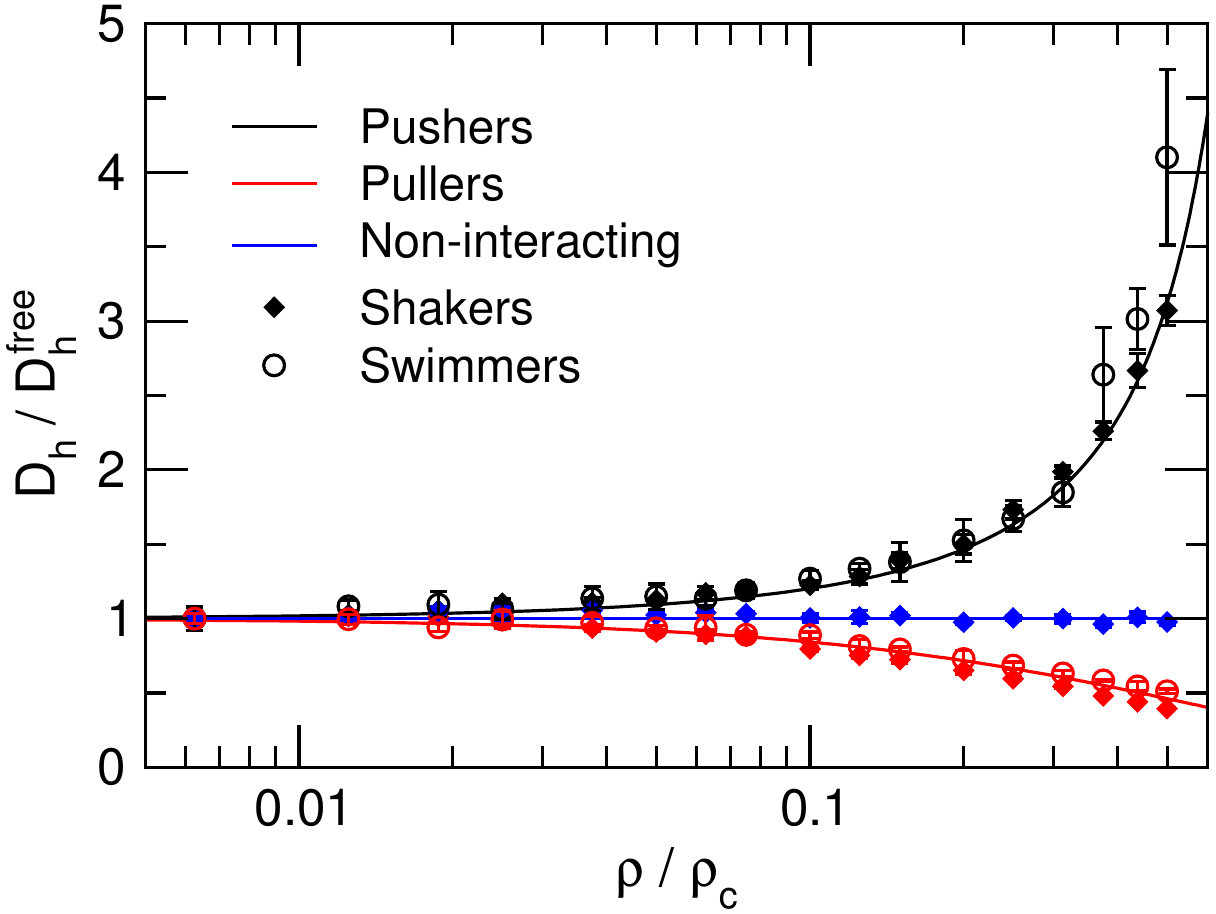}}
\caption{Diffusivity $D_h$, normalised by its value $D_h^{\mathrm{free}}$ in the non-interacting limit of passive tracers, as a function of the reduced swimmer density $\rho / \rho_c$. Symbols denote results from LB simulations of shakers or swimmers and solid lines show predictions of the kinetic theory for the same parameters as in Fig. \ref{fig:v2}. Error bars represent one standard deviation, estimated by averaging over four separate LB runs. The theoretical results were obtained by numerically solving the integrals before Eq. \eqref{eq:Dhapprox}.}
\label{fig:diff}
\end{center}
\end{figure}

In Fig. \ref{fig:diff}, we show that the enhanced diffusivity of tracers measured in LB simulations is perfectly described by the kinetic theory even close to the onset of turbulence. We further observe that $D_h$ deviates from $D_h^{\mathrm{free}}$ even for small densities of shakers, again highlighting the importance of correlations. Moreover, Eq. \eqref{eq:Dhapprox} correctly predicts how $D_h$ depends differently on $\rho$ for pushers and pullers -- an effect that has only briefly been discussed in the literature \cite{Underhill2008PRL,Ishikawa2010,kurtuldu2011,Subramanian2015JFM,Saintillan2011Interface}.

An approach previously used by several authors~\cite{childress2011,mino2011,Poon2013PRE,Yeomans2013PRL,kasyap2014hydrodynamic,morozov2014enhanced} to predict $D_h$ considers tracer displacements due to scattering from a single swimmer. This leads to $D_h \simeq D_h^{\mathrm{free}}\sim \rho \kappa_n^2$, and reflects the pusher-puller symmetry upon time-reversal in the Stokes equation. The extension of this argument to scattering by any finite number of swimmers presents a conceptual problem: the tracer displacement due to a scattering event by a collection of pushers can also be obtained in a suspension of pullers whose initial positions are set equal to the final positions of the pushers, with their orientations reversed. This argument thus suggests that $D_h$ should be equal for pushers and pullers at all densities, at odds with the numerical data in Fig. \ref{fig:diff}. The caveat in this argument is that it assumes a uniform sampling of initial conditions for the swimmers, while correlations between them will make some configurations more probable. When taken into account properly, these correlations break the pusher-puller symmetry even at moderate densities, as in Eq. \eqref{eq:Dhapprox}.

\emph{Conclusions.} In this Letter, we have presented results from a novel kinetic theory and unprecedently large particle-resolved simulations of microswimmer suspensions, describing quantitatively the fluctuations and correlations that arise at intermediate swimmer density. We have numerically shown that 
the collective motion in swimmer suspensions is the result of their rotational dynamics in the flow created by other swimmers, while self-propulsion and spatial correlations play subdominant roles. We calculated the fluid velocity fluctuations and the enhanced diffusivity of tracer particles, and found significant deviations from the mean-field predictions even at moderate swimmer densities. We demonstrated that swimmer-swimmer correlations are responsible for these deviations and should thus be taken into account well below the onset of bacterial turbulence. Understanding such correlations is a prerequisite for a deeper understanding of the turbulent state itself, in particular with regards to the presence or absence of a finite, characteristic length-scale in collective motion of microswimmers \cite{dombrowski2004self,sokolov2007concentration,Clement2014,Creppy2015,Sokolov2012,Dunkel2013PRL,Wensink2012PNAS}. 

\emph{Acknowledgements.} Discussions with Mike Cates and Joost de Graaf are kindly acknowledged. JS is financed by a grant from the Swedish Research Council (2015-05449), CN by EPSRC grant EP/J007404, RWN by Intel through EPCC's Intel Parallel Computing Centre, and AM by EPSRC through grant number EP/I004262/1.
CN acknowledges the hospitality provided by DAMTP, University of Cambridge while most of this work was being done and the support of an Aide Investissements d'Avenir du LabEx PALM (ANR-10-LABX-0039-PALM). Research outputs generated through the EPSRC grant EP/I004262/1 can be found at http://dx.doi.org/10.7488/ds/1703.

\emph{Author contributions.} JS and CN contributed equally to this work. 

\bibliographystyle{apsrev4-1}
\bibliography{active-matter}

\end{document}